\documentclass[prd,aps,nofootinbib,onecolumn,showpacs,floats,letterpaper,floatfix,groupedaddress,eqsecnum]{revtex4}

\usepackage{dcolumn,epsfig}
\usepackage{amssymb,amsmath}

\def\be{\begin{equation}}
\def\ee{\end{equation}}
\def\beq{\begin{eqnarray}}
\def\eeq{\end{eqnarray}}

\begin{document}

\title{Instability of hyper-compact Kerr-like objects}

\author{Vitor Cardoso} \email{vcardoso@fisica.ist.utl.pt}
\affiliation{Centro Multidisciplinar de Astrof\'{\i}sica - CENTRA, Dept. de F\'{\i}sica, Instituto
Superior T\'ecnico, Av. Rovisco Pais 1, 1049-001 Lisboa, Portugal
\\and Department of Physics and Astronomy, The University of Mississippi, University, MS 38677-1848, USA}

\author{Paolo Pani} \email{paolo.pani@ca.infn.it}
\affiliation{Dipartimento di Fisica, Universit\`a di Cagliari, and
INFN sezione di Cagliari, Cittadella Universitaria 09042 Monserrato,
Italy}

\author{Mariano Cadoni} \email{mariano.cadoni@ca.infn.it}
\affiliation{Dipartimento di Fisica, Universit\`a di Cagliari, and
INFN sezione di Cagliari, Cittadella Universitaria 09042 Monserrato,
Italy}

\author{Marco Cavagli\`a} \email{cavaglia@phy.olemiss.edu}
\affiliation{Department of Physics and Astronomy, The University of Mississippi, University, MS 38677-1848,
USA}

\date{\today}

\begin{abstract}
Viable alternatives to astrophysical black holes include
hyper-compact objects without horizon, such as gravastars, boson
stars, wormholes and superspinars. The authors have recently shown
that typical rapidly-spinning gravastars and boson stars develop a
strong instability. That analysis is extended in this paper to a
wide class of horizonless objects with approximate Kerr-like
geometry. A detailed investigation of wormholes and superspinars
is presented, using plausible models and mirror boundary
conditions at the surface. Like gravastars and boson stars, these
objects are unstable with very short instability timescales. This
result strengthens previous conclusions that observed
hyper-compact astrophysical objects with large rotation are likely
to be black holes.
\end{abstract}


\maketitle
\section{Introduction}
Astrophysical Black Holes (BHs) are believed to be common objects in galaxies. Their mass is
expected to span many orders of magnitude, from a fraction of the solar mass (primordial BHs in the
galactic halo) to few solar masses (stellar BHs in the galactic plane) up to several billions of
solar masses (supermassive BHs in galactic centers). Their angular momentum should be close to the
extremal limit due to accretion and mergers \cite{Gammie:2003qi, Merritt:2004gc}. For example, if
quasars are powered by supermassive BHs, astrophysical observations suggest that they should be
rotating near the Kerr bound \cite{Wang:2006bz}.

Unquestionable observational evidence of the existence of BHs is still lacking \cite{Narayan:2005ie,
Abramowicz:2002vt, Lasota:2006jh}. Current astrophysical data cannot rule out ``BH forgeries'', i.e.\
hyper-compact objects with redshift and geodesics similar to those of BHs, but lacking an event horizon.
Several models of hyper-compact objects with these characteristics have been known in the literature for
some time. Among these models, gravastars \cite{Chapline:2000en,Mazur:2001fv} and boson stars
\cite{bosonstars, Berti:2006qt} have been proposed as the most viable alternatives to astrophysical BHs.
The authors recently showed that rapidly spinning gravastars and boson stars may develop a strong
ergoregion instability \cite{Cardoso:2007az}. Their typical instability timescales are of order of $0.1$
seconds to 1 week for objects with mass $M = 1 - 10^6 M_{\odot}$. Therefore, observed astrophysical
hyper-compact objects are likely not to be gravastars nor boson stars.

The purpose of this paper is to compute the ergoregion instability
for other horizonless, Kerr-like hyper-compact objects: wormholes
and superspinars. Wormholes can be objects even simpler than BHs
\cite{Morris:1988cz, visserbook, Lemos:2003jb}. They are
infinitesimal variations of the Schwarzschild space-time which may
be indistinguishable from BHs \cite{Damour:2007ap}. In a string
theory context, the fuzzball model replaces BHs by horizonless
structures \cite{Mathur:2005zp}. The BH-like geometry emerges in a
coarse-grained description which ``averages'' over horizonless
geometries and produces an effective horizon at a radius where the
individual microstate geometries start to differ. Superspinars are
solutions of the gravitational field equations that violate the
Kerr bound.  These geometries could be created by high energy
corrections to Einstein gravity such as those present in
string-inspired models \cite{Gimon:2007ur, Matsas:2007bj}.
Superspinars are expected to have compactness of the order of
extremal rotating Kerr BHs and to exist in any mass range.

A rigorous analysis of the ergoregion instability for these models
is a non-trivial task; known wormhole solutions are special
non-vacuum solutions of the gravitational field equations. Thus
their investigation requires a case-by-case analysis of the
stress-energy tensor. Exact solutions of four-dimensional
superspinars are not known. To overcome these difficulties, the
following analysis will focus on a simple model which captures the
essential features of most Kerr-like horizonless hyper-compact
objects. Superspinars and rotating wormholes will be modeled by
the exterior Kerr metric down to their surface, where Dirichlet
boundary conditions are imposed. This problem is very similar to
Press and Teukolsky's ``BH bomb'' \cite{bhbombPress,
Cardoso:2004nk}, i.e. a rotating BH surrounded by a perfectly
reflecting mirror with its horizon replaced by a reflecting
surface. These boundary conditions are perfect mirror conditions
and require a reflection coefficient $R=1$. In a more realistic
model $R<1$ and a certain transmittance $T=1-R$ should be taken
into account, which will in principle decrease the strength of the
ergoregion instability. We argue that the qualitative behavior of
the instability is the same as long as the reflection or the
superradiant amplification are large enough. Letting $\rho$ be a
superradiant factor, one expects an ergoregion instability to
develop whenever $\rho (1-T)>1$. In the perfect mirror limit $T=0$
and the superradiant condition is simply $\rho>1$. 
The general case can be handled using both the analytical
and numerical techniques presented here. 

In Section \ref{sec:superspinars} we introduce the class of
objects we will deal with in this work. They are general
approximations to superspinars and wormholes with the basic key
features retained. In Section \ref{sec:wh} we show how to solve
for the instability analytically in two different regimes. The
details of these computations are left for Appendices
\ref{app:slow rotation} and \ref{app:extremal}. These
approximations are compared with numerical results in Section
\ref{sec:num}, where we also show that another kind of instability
sets in for general naked singularities. This ``algebraic''
instability can be computed algebraically in the Kerr geometry. We
close with a brief discussion of our results.

\section{\label{sec:superspinars} Superspinars and Kerr-like wormholes}
A superspinar of mass $M$ and angular momentum $J=aM$ can be modeled by the Kerr geometry
\cite{Gimon:2007ur}
\be
ds_{\rm Kerr}^2=-\left(1-\frac{2Mr}{\Sigma}\right)dt^2+\frac{\Sigma}{\Delta}dr^2 +
\left[(r^2+a^2)\sin^2\theta +\frac{2Mr}{\Sigma}a^2\sin^4\theta \right]d\phi^2-\frac{4Mr}{\Sigma}a\sin^2\theta d\phi dt
+{\Sigma}d\theta^2\,,
\label{kerrmetric}
\ee
where
\be \Sigma=r^2+a^2\cos^2\theta\,,\,\,\, \Delta=r^2+a^2-2M r\,.\ee
Unlike Kerr BHs, superspinars have $a>M$ and no horizon. Since the
domain of interest is $-\infty<r<+\infty$, the space-time posesses
naked singularities and closed timelike curves in regions where
$g_{\phi\phi}<0$ (see the monograph by Chandrasekhar
\cite{Chandraspecial}). High energy modifications in the vicinity
of the singularity are also expected. Following
Ref.~\cite{Gimon:2007ur}, a small region around the origin is
excised or assumed to be modified by, say, stringy corrections.
The most popular excision method uses domain walls formed by
supertubes \cite{Gimon:2007ur,Drukker:2004zm}. Kerr-like wormholes
are described by metrics of the form
\be ds^{2}_{\rm wormhole}=ds^2_{\rm Kerr}+\delta g_{ab}dx^{a}dx^b\,,
\label{worm}
\ee
where $\delta g_{ab}$ is infinitesimal. In general,
Eq.~(\ref{worm}) describes an horizonless object with excision at
some small distance of order $\epsilon$ from the would-be horizon.
(See Ref.\ \cite{Damour:2007ap} for details on nonrotating
wormholes). A detailed consideration of rotating stationary
wormholes, throat location and conditions on the metric is given
by Teo \cite{Teo:1998dp} to which we refer for further details.
Here, we simply assume these conditions are satisfied.
 Wormholes require exotic matter and/or divergent
stress tensors, thus some ultra-stiff matter is assumed close to
the would-be horizon. In the following, both superspinars and
wormholes will be modeled by the Kerr metric with a rigid ``wall''
at finite Boyer-Lindquist radius $r_0$, which excludes the
pathological region. We will consider both $a/M<1$ and $a/M>1$.

If the background geometries of superspinars and wormholes are sufficiently close to the Kerr
geometry, their dynamical perturbations are determined by the equations of perturbed Kerr BHs. The
proof is straightforward. Consider a minimally-coupled scalar field $\Phi$ propagating on a
space-time with metric
\be
g_{ab}=g^0_{ab}+\delta g_{ab}\,,
\ee
where $\delta g_{ab}\ll g_{ab}$. At first order in $\delta g_{ab}$, the Klein-Gordon equation
reads
\be
\partial_a\left [\sqrt{-g}g^{ab}\partial_b\Phi\left (1+\frac{1}{2}g^{cd}\delta g_{cd}\right)\right]+
\partial_a\left [\sqrt{-g}\delta g^{ab}\partial_b\Phi\left (1+\frac{1}{2}g^{cd}\delta g_{cd}
\right)\right]=0\,.
\label{kgexpanded}
\ee
If $g^{cd}\delta g_{cd}\ll 1$, Eq.\ (\ref{kgexpanded}) is identical to the equation of a scalar field in
the Kerr geometry. This result also generalizes to the Maxwell field. Gravitational perturbations can be
handled as in the Kerr geometry only if they are larger than $\delta g_{ab}$ at any time.
\section{Instability analysis: analytic results} \label{sec:wh}
The instability of superspinars and wormholes is studied by considering Kerr geometries with
arbitrary rotation parameter $a$ and a ``mirror'' at some Boyer-Lindquist radius $r_0$. Using the
Kinnersley tetrad and Boyer-Lindquist coordinates, it is possible to separate the angular variables
from the radial ones, decoupling all quantities. Small perturbations of a spin-$s$
field are reduced to the radial and angular master equations \cite{teukolsky}
\be
\Delta^{-s}\frac{d}{dr}\left(\Delta^{s+1}\frac{dR_{l
m}}{dr}\right)+ \left[\frac{K^{2}-2is(r-M)K}{\Delta}+4is\omega r -\lambda\right]R_{l m}=0\,,
\label{wave eq separated general}
\ee
\be
\left[(1-x^2){}_sS_{l m,x} \right]_{,x}+
\left[(a\omega x)^2-2a\omega sx+s+{}_{s}A_{l m}-\frac{(m+sx)^2}{1-x^2}\right]{}_sS_{l m}=0\,,
\label{angularwaveeq}
\ee
where $x\equiv\cos\theta$, $\Delta=r^2-2Mr+a^2$, $K=(r^2+a^2)\omega-am$ and the separation constants
$\lambda$ and ${}_s A_{l m}$ are related by
\be
\lambda \equiv {}_s A_{l m}+a^2\omega^2-2am\omega\,.
\label{sAlm}
\ee
If $a\leqslant M$, the space-time possesses one or two horizons located at $r_{\pm}=M\pm\sqrt{M^2-a^2}$.
Equations (\ref{wave eq separated general})-(\ref{angularwaveeq}) can be analytically solved in the
slowly-rotating and low-frequency regime, $\omega M\ll 1$ \cite{staro1,Unruh:1976fm,Cardoso:2004nk}, and in
the rapidly-spinning regime, where $r_+\sim r_-$ and $\omega \sim m \Omega_h$, $\Omega_h \equiv a/(2Mr_+)$
being the angular velocity at the horizon \cite{staro1}.
\subsection{\label{sec:slow rotation} Slowly rotating objects and low frequencies}
%
\begin{figure*}[h]
\begin{tabular}{ll}
\epsfig{file=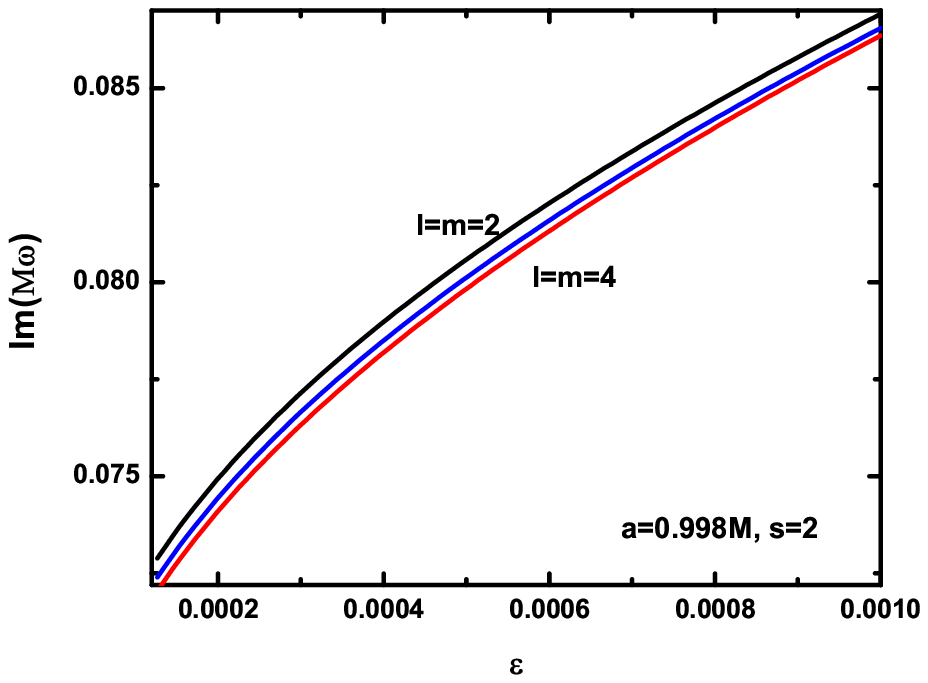,height=140pt,angle=0} &
\epsfig{file=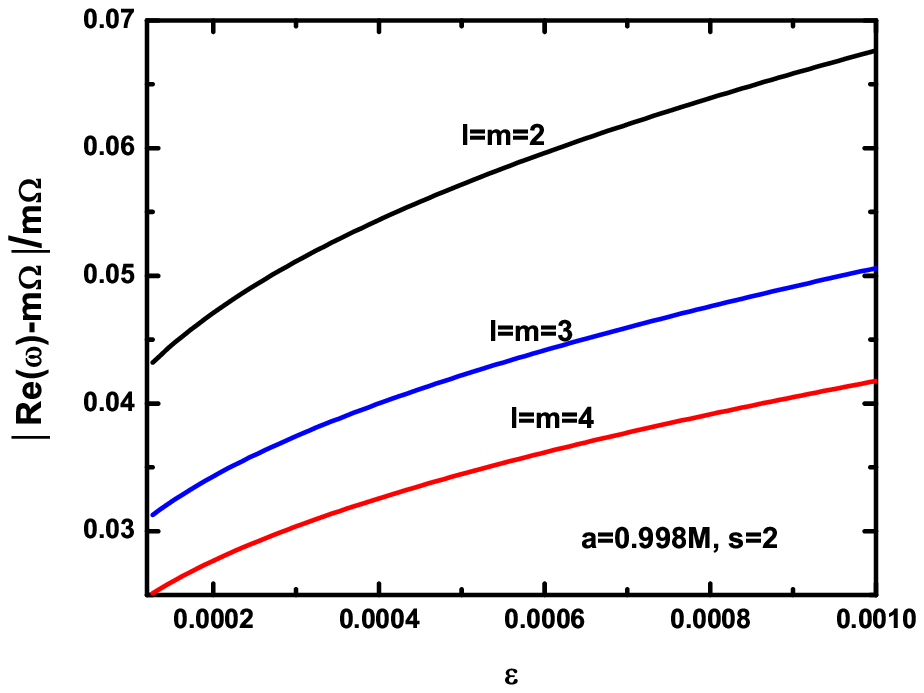,height=140pt,angle=0}
\end{tabular}
\caption{Imaginary and real parts of the characteristic gravitational frequencies for an object with
$a=0.998M$, according to the analytic calculation for rapidly-spinning objects. The mirror location
is at $r_0=(1+\epsilon)r_+$. The real part is approximately constant and close to $m\Omega$, in agreement
with the assumptions used in the analytic approach.}
\label{fig:extremal scalar}
\end{figure*}
The instability timescales for small rotation and low frequencies can be computed by approximating the
Teukolsky equation near the horizon and then matching its solution to the solution near infinity.  The
spheroidal wavefunctions (\ref{angularwaveeq}) reduce to the spin-weighted spherical harmonics with
eigenvalues ${}_sA_{l m}=l(l+1)-s(s+1)$. Matching the inner solution to the outer solution, the frequency
$\omega$ is (see Appendix \ref{app:slow rotation} for details)
\be
z_0^{-s-2i\varpi}=\prod_{k=1}^{l}\left(\frac{k+s+2i\varpi}{k-s-2i\varpi}\right)
\left[\frac{1-i(\omega(r_+ -r_-))^{2l+1}
L_{s}(s+2i\varpi)\prod_{k=1}^{l}(k^2+4\varpi^{2}-s^{2}-4is\varpi)}
{1+i(\omega(r_+ -r_-))^{2l+1}
L_{s}(s+2i\varpi)\prod_{k=1}^{l}(k^2+4\varpi^{2}-s^{2}-4is\varpi)}\right]\,,
\label{relation mirror zeros gen2}
\ee
where
\be L_s=2^{2l-1}(-1)^s\Gamma(1+l-s)\Gamma(1+l+s)\left[\frac{\Gamma(l+1)}{\Gamma(2l+2)
\Gamma(2l+1)}\right]^{2}\,.
\label{Ls}
\ee
Equation (\ref{relation mirror zeros gen2}) can be solved numerically for the characteristic
values of $\omega$. Care must be exercised to ensure that the solutions are consistent with
the approximation scheme, i.e., $M\omega\ll 1$, $a\ll M$ and $z_0\sim 0$. The instability
timescale for scalar perturbations is
\be
\frac{\tau}{M}= 2\left[\frac{1-(a/\sqrt{2}M)^2+
\sqrt{1-(a/M)^2}}{\sqrt{1-(a/M)^2}}\right]\cdot\log^{-1}
\left[\frac{1+\gamma(\omega_{n,m}-m\Omega)}{1-\gamma(\omega_{n,m}-m\Omega)}\right]\,.
\label{relation time scale unit2}
\ee
\begin{table}[ht]
\centering
\caption{\label{tab:approxinstab} Approximate solution for a slowly rotating object with
$a=0.3M$ and mirror position at $\epsilon=10^{-5}$. The two columns give the real part of the
frequency $Re(\omega)$ and the growth time $\tau$ for different $l=m$ values of the scalar
field, respectively. The mode $l=m=1$ is absent.}
\begin{tabular}{|c||cc|}
%
\hline \multicolumn{1}{|c}{} & \multicolumn{2}{c|}{$a/M=0.3$, $\epsilon=10^{-5}$}\\
\hline

$l=m$  & $Re(\omega)M$                    &$\tau/M$\\
2    & $0.02$         &$1.8\times 10^{13}$    \\
3    & $0.09$        &$5.8\times 10^{13}$    \\
4    & $0.17$        &$8.1\times 10^{15}$    \\
\hline\hline
\end{tabular}
\end{table}
Table \ref{tab:approxinstab} shows the results for $s=0$ and
different values of $l=m$. The scalar instability timescale is
much smaller than the Hawking evaporation timescale
\cite{hawking_evap}, $\tau_{\rm evap}\approx 10^{71}{\left
(M/M_\odot\right)}^{3}{\rm sec}$, and increases with $l=m$.
Gravitational instability timescales of fast rotating objects are
expected to be larger by several orders of magnitude.
\subsection{\label{sec:fast rotation} Fast-rotating objects}
Defining $x=(r-r_+)/r_+$, $\sigma=(r_+-r_-)/r_+$, $\tau=M(\omega-m\Omega)$ and $\omega'\rightarrow
r_{+}\omega$, the Teukolsky equation can be written as
\be
[2i\omega' x^2+x(4i\omega'-2(s+1))-(s+1)\sigma+4i\tau] R'(x)
+[2(2s+1)i\omega'(x+1)+\lambda]R(x)-x(x+\sigma) R''(x)=0\,.
\label{redef2}
\ee
Equation (\ref{redef2}) can be solved in the limit $a\rightarrow M$ and $\omega\rightarrow
m\Omega$ by following the procedure of the previous section. The relation between the position
of the mirror and the frequency of the wave is (see Appendix \ref{app:extremal} for details)
\be
(-z_0)^{-s-i\kappa}=\frac{R_1+\rho R_3(-2i\omega\sigma)^{2i\delta}}{R_2+\rho R_4
(-2i\omega\sigma)^{2i\delta}}\,.
\label{extremal zero equation2}
\ee
Numerical solutions of the above equation for a star with $a=0.998M$ are shown in
Fig.~\ref{fig:extremal scalar}. The real part of the characteristic frequency is always close to
$m\Omega$ and ${\rm Im}(\omega)\ll {\rm Re}(\omega)$. Thus the results are consistent with the
initial assumptions. The instability timescale for gravitational perturbations is about five orders
of magnitude smaller than the instability timescale for scalar perturbations.
\section{Instability analysis: numerical results \label{sec:num}}
The oscillation frequencies of the modes can be found from the  canonical form of Eq.~(\ref{wave eq
separated general})
\be
\frac{d^2Y}{dr_*^2}+VY=0\,,
\label{teu canonical}
\ee
where
\be
Y=\Delta^{s/2}(r^2+a^2)^{1/2}R\,,\qquad
V=\frac{K^2-2is(r-M)K+\Delta(4ir\omega
s-\lambda)}{(r^2+a^2)^{2}}-G^2-\frac{dG}{dr_*}\,,
\ee
and $K=(r^2+a^2)\omega-am$, $G=s(r-M)/(r^2+a^2)+r\Delta(r^2+a^2)^{-2}$. The separation constant
$\lambda$ is related to the eigenvalues of the angular equation by Eq.~(\ref{sAlm}). The eigenvalues
${}_sA_{lm}$ are expanded in power series of $a\omega$ as \cite{Berti:2005gp}
\be {}_sA_{lm}=\sum_{k=0}f^{(k)}_{slm}(a\omega)^{k}\,.
\label{rel:eigexpans} \ee
Terms up to order $(a\omega)^2$ are included in the calculation.  Absence of ingoing waves at infinity
implies
\be
Y\sim r^{-s}e^{i\omega r_*}\,.
\label{asymp sol}
\ee
Numerical results are obtained by integrating Eq.~(\ref{teu canonical}) inward from a large distance
$r_{\infty}$. The integration is performed with the Runge-Kutta method with fixed $\omega$ starting
at $M r_\infty = 400$, where the asymptotic behavior (\ref{asymp sol}) is imposed. (Choosing a
different initial point does not affect the final results.) The numerical integration is stopped at
the radius of the mirror $r_0$, where the value of the field $Y(\omega,r_0)$ is extracted. The
integration is repeated for different values of $\omega$ until $Y(\omega,r_0)=0$ is obtained with the
desired precision. If $Y(\omega,r_0)$ vanishes, the field satisfies the boundary condition for
perfect reflection and $\omega=\omega_0$ is the oscillation frequency of the mode.
\subsection{Objects with $a<M$}
\begin{figure*}[h]
\begin{tabular}{cc}
\epsfig{file=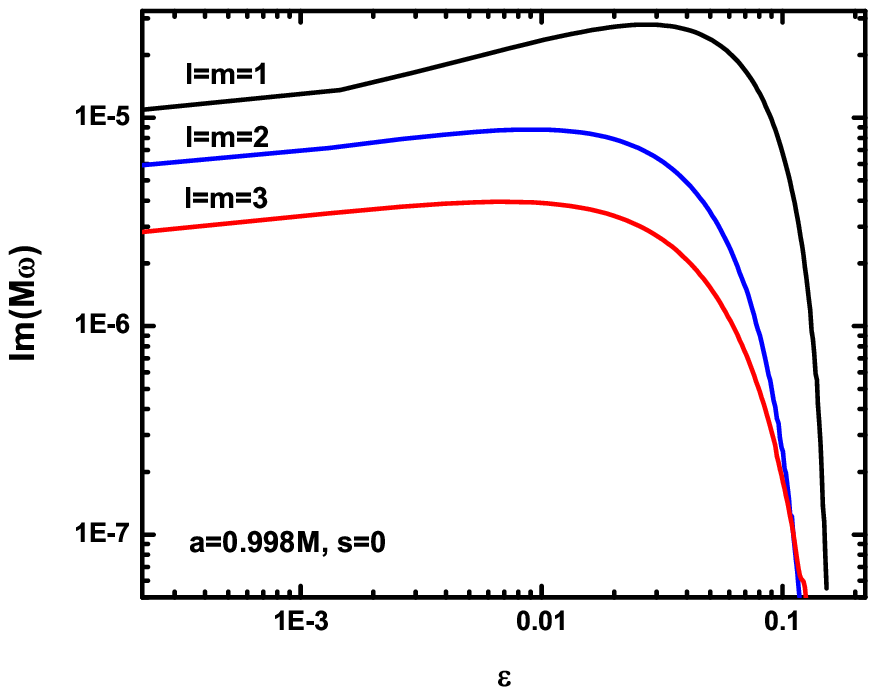,height=140pt,angle=0}&
\epsfig{file=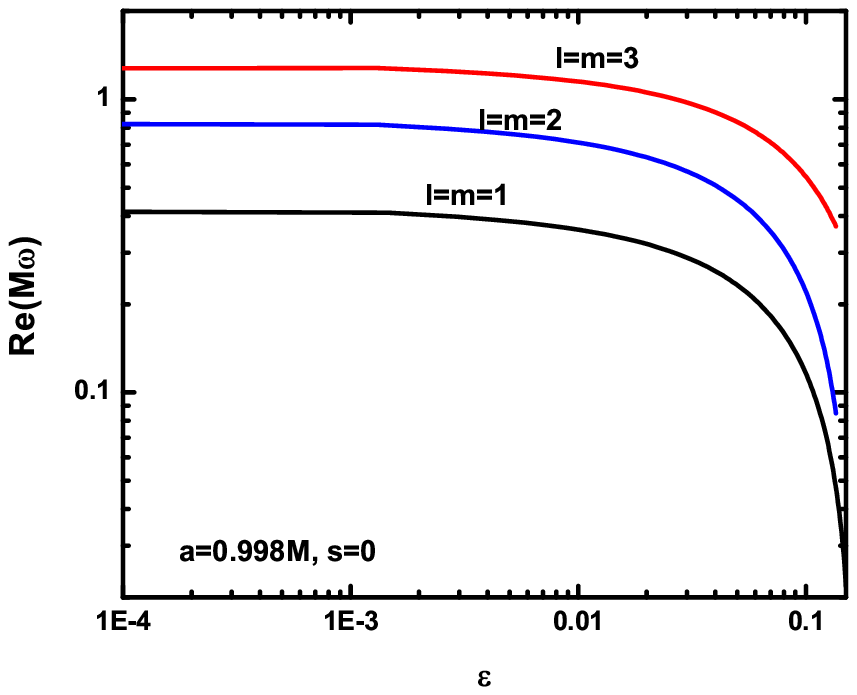,height=140pt,angle=0}\\
\epsfig{file=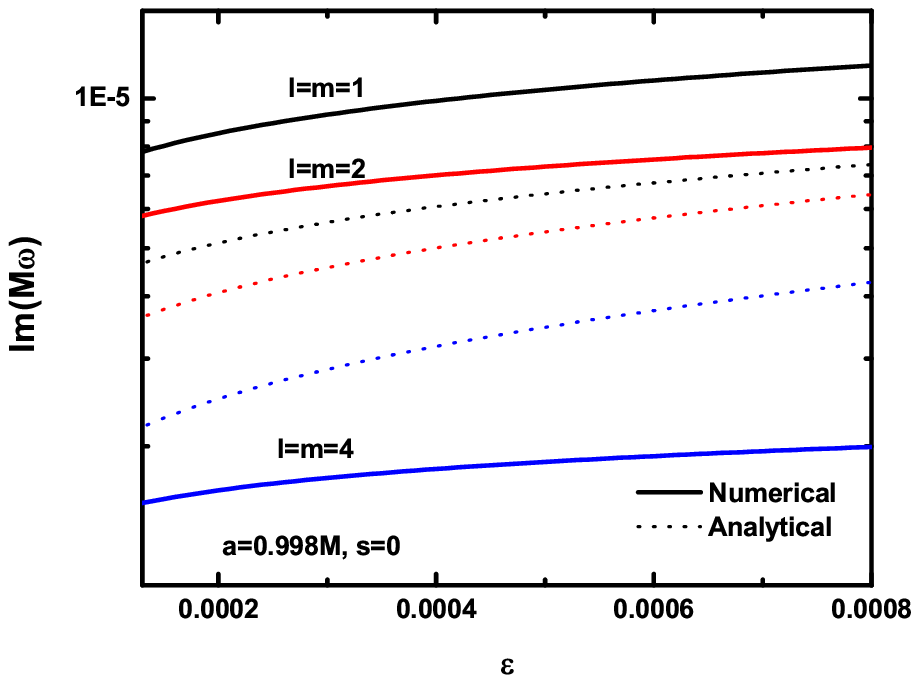,height=140pt,angle=0}&
\epsfig{file=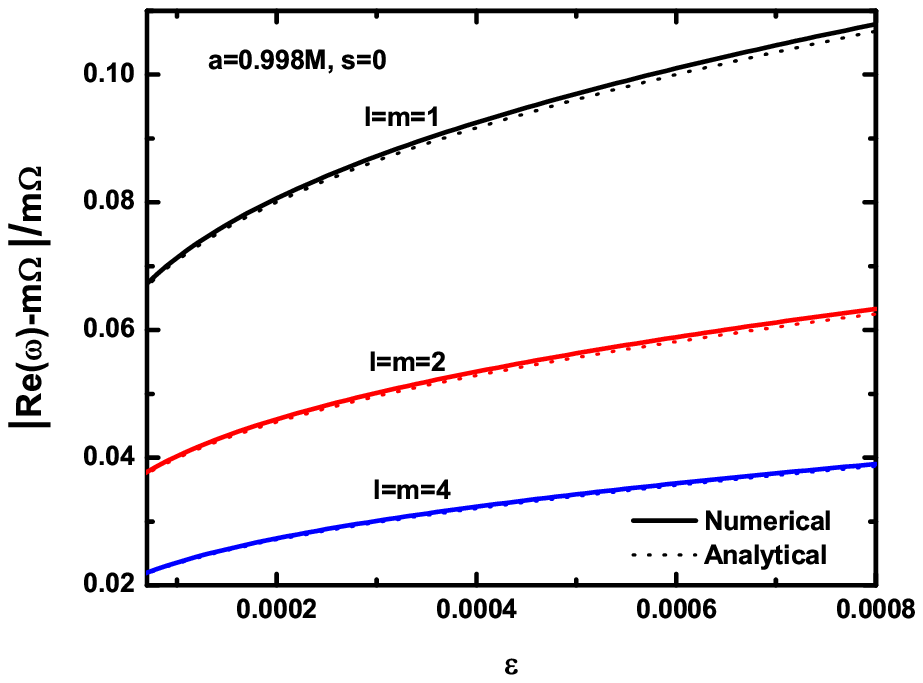,height=140pt,angle=0}
\end{tabular}
\caption{Details of the instability for scalar perturbations, from a numerical solution of Teukolsky
equation. Top panels: Numerical results for the imaginary part (left panel) and real part (right panel) of
the fundamental mode frequency vs.\ the mirror position $r_0=r_+(1+\epsilon)$ for different $l=m$ values.
The angular momentum is $a=0.998M$. The instability grows monotonically with $l=m$. Bottom panels:
Zoomed-in version of the upper panels, where numerical results (solid lines) are compared to the analytic
solutions in the near-extremal regime (\ref{extremal zero equation2}) (dotted lines). The agreement between
numerical and analytic values of the resonant frequency is remarkable. The analytic results for the
imaginary part (bottom left panel) agree with the numerical results within a factor $\sim 3$.}
\label{fig:instscalar0.499ls0}
\end{figure*}

The regime $a<M$ requires a surface or mirror at $r_0=r_+(1+\epsilon)>r_+$. Typical results for scalar
perturbations of objects with $a<M$ are summarized in Table \ref{tab:inststarbomb} and Fig.\
\ref{fig:instscalar0.499ls0}. The top panels of Fig.\ \ref{fig:instscalar0.499ls0} show the imaginary and
real parts of the fundamental mode frequency for $a=0.998M$ vs.\ the mirror position $r_0=r_+(1+\epsilon)$
for different $l=m$ values, respectively.  The instability is weaker for larger $m$. This result holds also
for $l\ne m$, although it is not shown in the plots. Comparisons with the analytic results in the
near-extremal regime are shown in the bottom panels. The numerical integrations indicate that ${\rm
Re}(\omega)\sim m\Omega$, in agreement with the analytic results of Sect.~\ref{sec:fast rotation}. The
instability timescales are consistent with the analytic results within a factor $\sim 3$. The minimum
instability timescale is of order $\tau \sim 10^5M$ for a wide range of mirror locations.

\begin{table}[ht]
\caption{\label{tab:inststarbomb} Characteristic frequencies and
instability timescales for a Kerr-like object with $a=0.998M$. The
mirror is located at $\epsilon=0.1$. Results for scalar and
gravitational perturbations are shown for different $l=m$ values.}
\begin{tabular}{|c||cc|}
%
\hline \multicolumn{1}{|c}{} & \multicolumn{2}{c|}{ $({\rm Re}(\omega)M\,,{\rm Im}(\omega)M)$}\\
\hline
$l=m$  & $s=0$                    &$s=2$\\
1    & $(0.1120\,,0.6244\times 10^{-5})$          &$-$    \\
2    & $(0.4440\,,0.5373\times 10^{-5})$         &$ (0.4342\,,0.2900)$    \\
3    & $(0.7902\,,0.1928\times 10^{-5})$        &$(0.7803\,,0.2977)$    \\
4    & $(1.1436\,,0.5927\times 10^{-6})$        &$ (1.1336\,,0.3035)$    \\
\hline \hline
\end{tabular}
\end{table}

\begin{figure}[ht]
\begin{tabular}{cc}
\epsfig{file=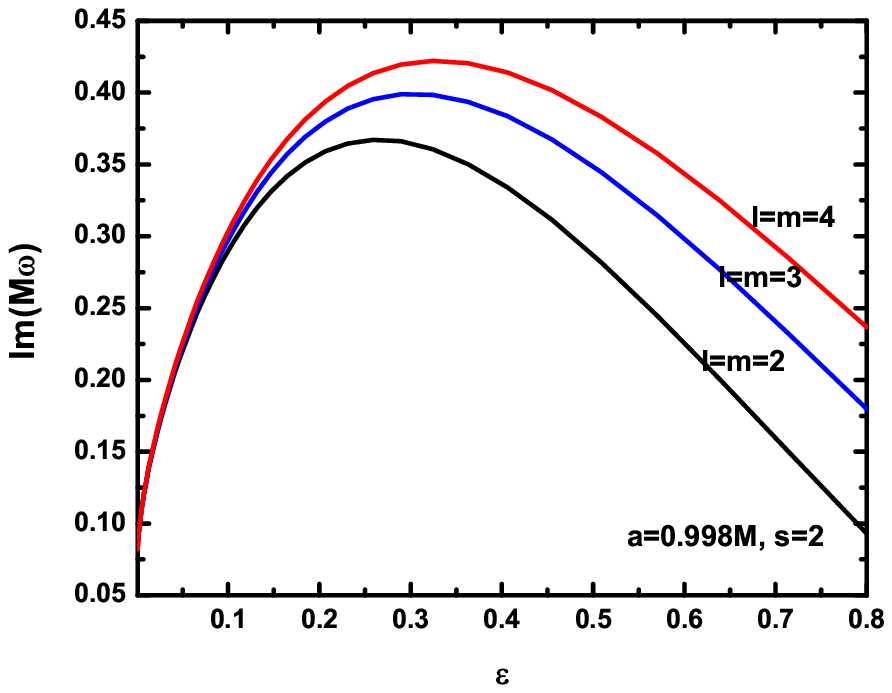,height=140pt,angle=0} &
\epsfig{file=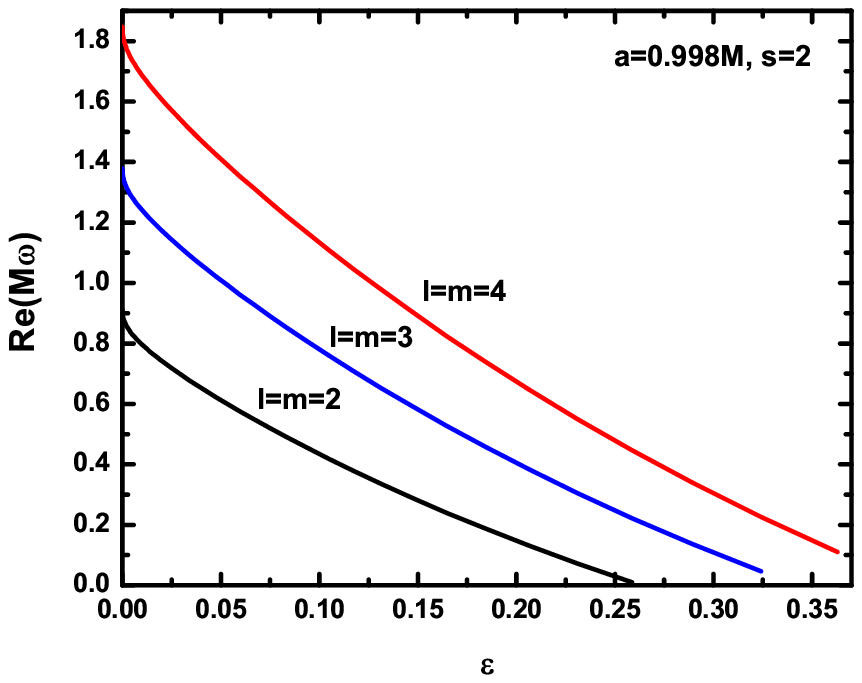,height=140pt,angle=0}\\
\epsfig{file=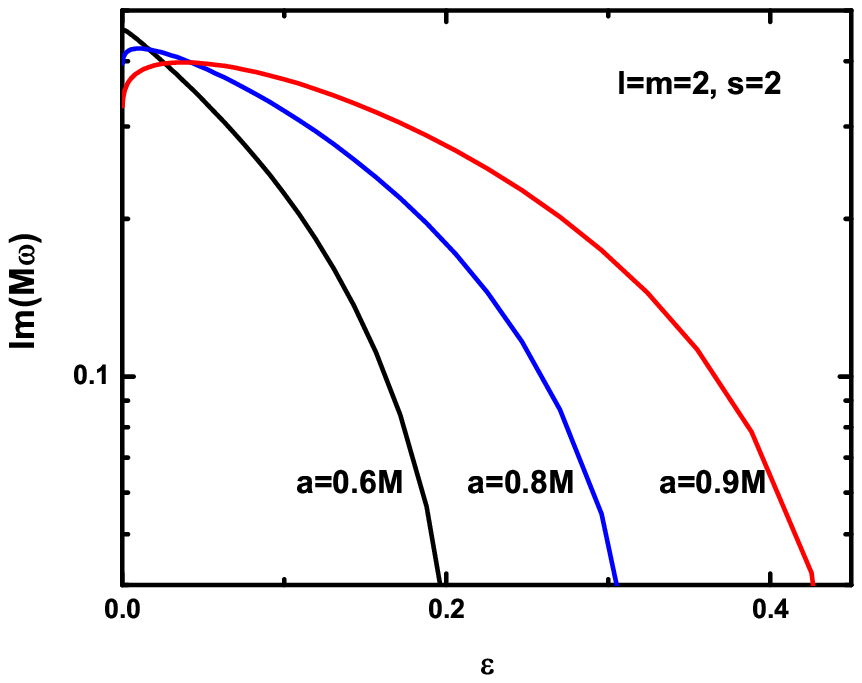,height=140pt,angle=0}&
\epsfig{file=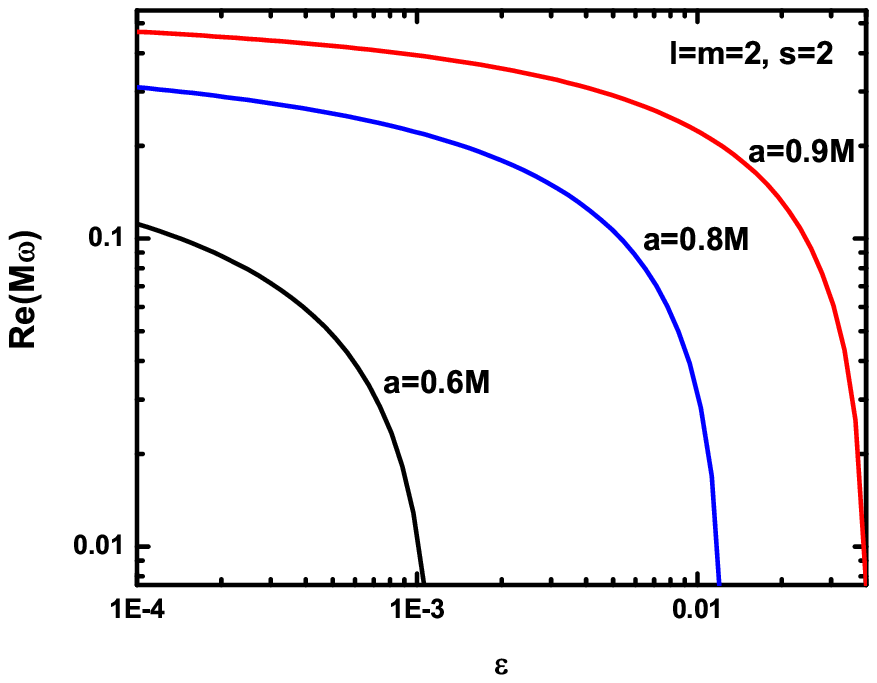,height=140pt,angle=0}
\end{tabular}
\caption{Details of the instability for gravitational perturbations, for different $l=m$ modes and $a/M=0.998$
(top panels) and for $l=m=2$ and different $a/M<1$.}
\label{fig:severalrotation}
\end{figure}

Figure \ref{fig:severalrotation} shows the results for gravitational perturbations. Instability timescales
are of the order of $\tau \sim 2-6M$. Thus gravitational perturbations lead to an instability about five
orders of magnitude stronger than the instability due to scalar perturbations (see Table
\ref{tab:inststarbomb}). Figure \ref{fig:severalrotation} shows that the ergoregion instability remains
relevant even for values of the angular momentum as low as $a=0.6M$.

Some features of these results are intriguing and deserve further study. For instance, the regime ${\rm
Re}(\omega)>m\Omega$ may also be unstable. Since the superradiant instability is confined to the
superradiant regime, this mode should describe a different kind of instability. Numerical results also show
that there is a mirror location which maximizes the instability at fixed $a$. It would be interesting to
explain the physical meaning of this location.
\subsection{Objects with $a>M$}
\begin{figure}[ht]
\begin{tabular}{cc}
\epsfig{file=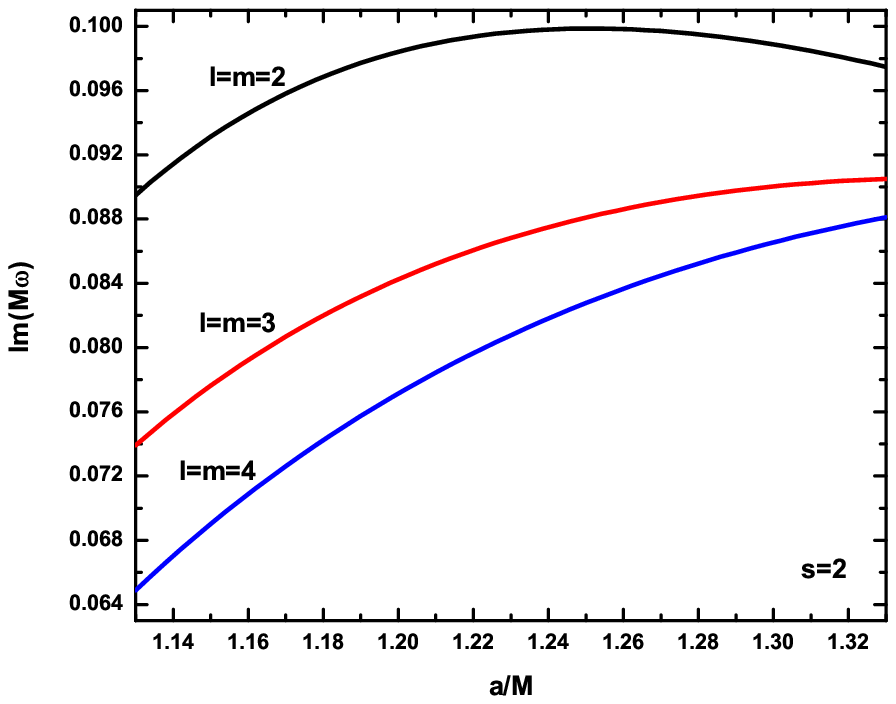,height=.37\textwidth,angle=0} &
\epsfig{file=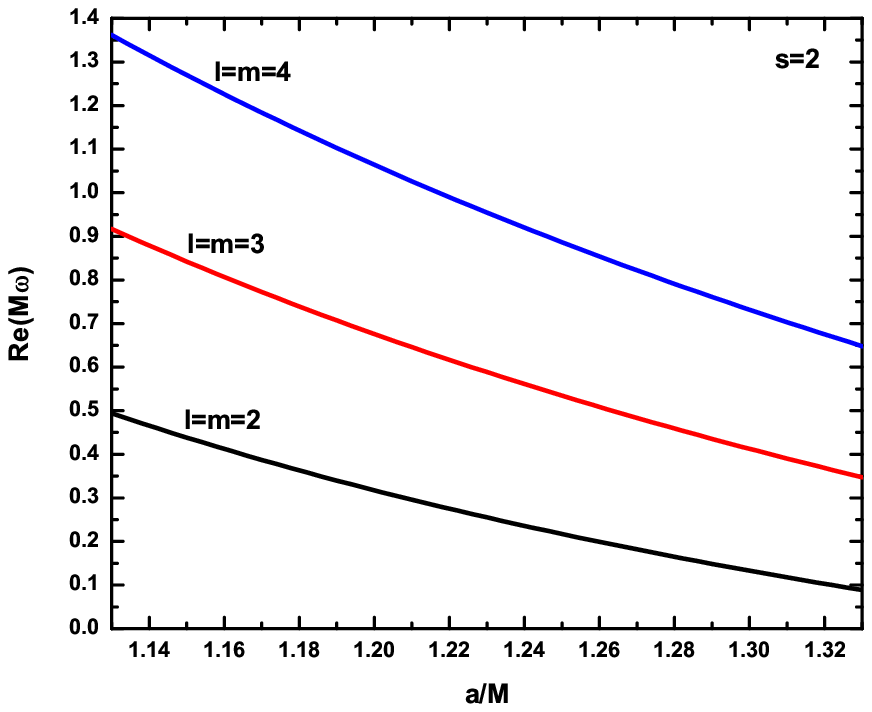,height=.35\textwidth,angle=0}
\end{tabular}
\caption{The fundamental $l=m=2,3,4$ modes of an object spinning above the Kerr bound as function of
rotation. The surface is located at $r_0/M=0.001$.}
\label{fig:super}
\end{figure}

Objects with $a>M$ could potentially describe superspinars.
Several arguments suggest that objects rotating above the Kerr
bound are unstable. Firstly, extremal Kerr BHs are marginally
stable. Thus the addition of extra rotation should lead to
instability. Secondly, fast-spinning objects usually take a
pancake-like form \cite{Emparan:2003sy} and are subject to the
Gregory-Laflamme instability \cite{Gregory:1993vy,Cardoso:2006ks}.
Finally, Kerr-like geometries, like naked singularities, seem to
be unstable against a certain class of gravitational perturbations
\cite{Gleiser:2006yz, Cardoso:2006bv, Dotti:2006gc} called
algebraically special perturbations \cite{Chandraspecial}. These
perturbations are described by modes with zero
Teukolsky-Starobinsky constant \cite{teukolsky}, and will be
discussed in more detail below. For objects with $a>M$ the surface
or mirror can be placed anywhere outside $r=0$. In general the
instability is as strong as in the $a<M$ regime. An example in
shown in Fig.\ \ref{fig:super}, which displays the resonant
frequencies for the instability of a surface at $r_0/M=0.001$.
This result seems to confirm other investigations suggesting that
ultra-compact objects rotating above the Kerr bound are unstable
\cite{Dotti:2008yr}.

\subsection{Algebraically special modes \label{as}}
Naked singularities are also characterized by a further kind of instability which is not related to the
presence of an ergoregion \cite{Gleiser:2006yz, Cardoso:2006bv, Dotti:2006gc}. This instability is
generated by modes with vanishing Teukolsky-Starobinsky constant \cite{teukolsky}. These  ``algebraically
special modes'' are regular for $r>0$ and may be relevant to superspinar geometries without pathological
regions. They can be expressed in analytic form. The solution of the Teukolsky equation is
\be
R_{l m}(r) = (A+Br +Cr^2 +Dr^3)\, e^{i\omega^{\rm AS}
(r_*-t)+im\phi}S_{l m}(\theta) \,,
\ee
where $A$, $B$, $C$, $D$ are constants and $S_{l m}$ are spin weight-2 spheroidal harmonics
\cite{Berti:2005gp}. These modes satisfy proper boundary conditions at infinity and are well behaved
for any $r>-\infty$. Although superspinars require particular boundary conditions at the excised
region, unstable algebraically special modes can be present. They can be computed with the continued
fraction method \cite{Leaver:1985ax,Onozawa:1996ux} and correspond to a zero of the
Teukolsky-Starobinsky constant squared:
\be
|C|^2= \lambda^2\left(\lambda+2\right
)^2-8\omega^2\lambda\left[\alpha^2(6+5\lambda)-12a^2\right]+144\omega^2\left(\alpha^4\omega^2+M^2\right)\,,
\ee
where $\lambda$ is defined in Eq.~(\ref{sAlm}) for $s=-2$ and $\alpha^2=a^2-ma/\omega$. (Note a
typographical error in Ref.~\cite{Onozawa:1996ux}.) The technique of Ref.~\cite{Leaver:1985ax} can
be used to evaluate the algebraically special modes at fixed $a$. In the range $0<a<M$, the modes
coincide with those of Ref.~\cite{Onozawa:1996ux}. Some results for $a>M$ are listed in Table
\ref{tab:TSconst}. The typical timescales are of the order of $10^{-6}$ seconds for a $1 M_\odot$
star and $1$ second for $M=10^6 M_{\odot}$.

\begin{table}
\begin{center}
\begin{tabular}{|c|c|c|}
\hline \multicolumn{1}{|c}{} & \multicolumn{2}{c|}{ $2M \omega$}\\
\hline
 $\frac{a}{M}$    & m=0     & m=2\\ \hline
 0.00 & (0, 4.0000)        & (0.0000, 4.0000)  \\ \hline
 0.10 & (0, 4.03107)       & (0.918857,3.66121)\\ \hline
 0.20 & (0, 4.13218)       & (1.36523,3.07054) \\ \hline
 0.30 & (0, 4.33511)       & (1.51483,2.57005) \\ \hline
 0.40 & (0, 4.74699)       & (1.53973,2.19077) \\ \hline
 0.50 & (0.715878,6.57057) & (1.51441,1.90331) \\ \hline
 0.60 & (2.47766,5.25615)  & (1.46928,1.68066) \\ \hline
 0.70 & (2.84578, 4.36141) & (1.41732,1.50403) \\ \hline
 0.80 & (2.93170, 3.71616) & (1.36428,1.36084) \\ \hline
 0.90 & (2.91142, 3.23046) & (1.31270,1.24256) \\ \hline
 1.00 & (2.84632, 2.85259) & (1.26369,1.14328) \\ \hline
 1.02 & (2.83051, 2.78693) & (1.25424,1.12531) \\ \hline
 1.04 & (2.81414, 2.72409) & (1.24490,1.10790) \\ \hline
 1.10 & (2.76243, 2.55079) & (1.21763,1.05878) \\ \hline
 1.20 & (2.67192, 2.30455) & (1.17458,0.98601) \\ \hline
 1.30 & (2.58073, 2.10007) & (1.13444,0.92268) \\ \hline
 1.40 & (2.49185, 1.92767) & (1.09701,0.86707) \\ \hline
\end{tabular}
\end{center}
\caption{\label{tab:TSconst} Algebraically special modes for various
values of the angular momentum.}
\end{table}
%
\section{Discussion\label{sec:discussion}}
This paper presented a general method for investigating the ergoregion instability of ultra-compact,
horizonless  Kerr-like objects. The essential features of these objects have been captured by a simple
model whose physical properties are largely independent from the dynamical details of the gravitational
system. The method has been applied to superspinars and rotating wormholes.  Numerical and analytic
results show that the ergoregion instability of these objects is extremely strong for any value of their
angular momentum, with timescales of order $10^{-5}$ seconds for a $1 M_\odot$ star and $10$ seconds for a
$M=10^6 M_{\odot}$ star. The above investigation confirms previous results for gravastars and boson stars
\cite{Cardoso:2007az}, namely that exotic objects without event horizon are likely to be ruled out as
viable candidates for astrophysical hyper-compact objects.
\section*{Acknowledgements}
The authors are grateful to Matteo Losito for a careful reading of
the manuscript. This work was partially funded by Funda\c c\~ao
para a Ci\^encia e Tecnologia (FCT) - Portugal through projects
PTDC/FIS/64175/2006 and POCI/FP/81915/2007. One of the authors
(MC) gratefully acknowledges the support of the National Science
Foundation through LIGO Research Support grant NSF PHY-0757937.

\appendix
\section{\label{app:slow rotation} Analytic solution in the low-frequency regime}
This appendix contains the analytic computation of the instability
of rotating objects, bounded by a hard wall. Massless and massive
scalar perturbations and general spin-$s$ perturbations are
considered.
\subsection{Massless scalar fields}
Following Ref.\ \cite{staro1,Unruh:1976fm,Cardoso:2004nk}, the space-time outside the star is divided in
a near region, $r-r_+ \ll 1/\omega$, and a far region, $r-r_+ \gg M$. The radial equation (\ref{wave eq
separated general}) is solved separately in each of these two regions with the assumptions that the Compton
wavelength of the scalar particle is much larger than the typical size of the object, $1/\omega \gg M$, and
$a\ll M$. These solutions are then matched in the overlapping region, where the condition $M \ll r-r_+ \ll
1/\omega$ is satisfied. The equation for the characteristic value $\omega$ is obtained by imposing suitable
boundary conditions at the boundaries. In the following, $r_+$ denotes the location of the ``would-be''
horizon. The location of the mirror or ultra-stiff wall is $r_0=r_+(1+\epsilon)$, $\epsilon\ll 1$.

In the far region, where the effects induced by the BH can be neglected, one can approximate $a\sim 0$, $M
\sim 0$ and $\Delta \sim r^2$. The radial wave equation reduces to the wave equation for a massless scalar
field of frequency $\omega$ and angular momentum $l$ in flat background
\be
\partial_r^2 (r R)+ \left[\omega^2-l(l+1)/r^2\right] (r R)=0\,.
\label{far wave eq}
\ee
The most general solution of this equation is the linear combination of Bessel functions
\cite{abramowitz}
\be
R=r^{-1/2}\left[\alpha J_{\,l+1/2}(\omega r)+\beta J_{-l-1/2}(\omega r)\right]\,.
\label{far field}
\ee
The behavior of Eq.~(\ref{far field}) for large $r$ and small $r$ are
\be
R \sim \sqrt{\frac{2}{\pi \omega}}\frac{1}{r}\left[
\alpha \sin(\omega r-l\pi/2)+ \beta \cos(\omega r+l\pi/2)\right]\,,\qquad
R \sim \alpha\, \frac{(\omega/2)^{l+1/2}}{\Gamma(l+3/2)}\: r^{l}
+ \beta\, \frac{(\omega/2)^{-l-1/2}}{\Gamma(-l+1/2)}\: r^{-l-1}\,,
\label{far field 2}
\ee
respectively. Absence of ingoing waves at infinity implies $\beta=-i\alpha e^{i\pi l}$. The radial wave
equation in the near region is
\beq
\Delta\partial_r \left ( \Delta\partial_r R \right )+ r_+^4
(\omega-m\Omega)^2\,R - l(l+1)\Delta\,R=0 \,.
 \label{near wave eq}
\eeq
Introducing the radial coordinate
\beq
z=\frac{r-r_+}{r-r_-}\,, \qquad 0\leq z \leq 1\,,
 \label{new radial coordinate}
\eeq
Eq.~(\ref{near wave eq}) can be rewritten as
\be
z(1\!-\!z)\partial_z^2 F+\left\{(1+i\, 2\varpi)-\left[
1+2(l+1)+ i\, 2\varpi\right]\,z \right\}
\partial_z F -\left [ (l+1)^2+ i \,2\varpi (l+1)\right ] F=0\,,
\label{hypergeometric equation}
\ee
where $R=z^{i \,\varpi} (1-z)^{l+1}F$ and
\be
\varpi \equiv(\omega-m\Omega)\frac{r_+^2}{r_+-r_-} \,.
 \label{superradiance factor}
\ee
Equation (\ref{hypergeometric equation}) is a standard hypergeometric equation. Its most general
solution is
\be
R=A\, z^{-i\,\varpi}(1-z)^{l+1}
F(a-c+1,b-c+1,2-c,z)+B\,z^{i\,\varpi}(1-z)^{l+1} F(a,b,c,z) \,,
\label{hypergeometric solution}
\ee
where $a=1+l+i\,2\varpi$, $b=l+1$ and $c=1+ i\,2\varpi$. Near the wall, where $r\sim r_+$, Eq.~(\ref{hypergeometric solution}) reads
\be
R \approx A\, z^{-i\,\varpi}+B\,z^{i\,\varpi}=A\,e^{-i(\omega-m\Omega)\frac{r_+}{2M}r_*}+
B\,e^{i(\omega-m\Omega)\frac{r_+}{2M}r_*}\,,
\label{relation near-r near-sol}
\ee
where $r_*$ is the tortoise coordinate
\be
r_*=\int \frac{r^2+a^2}{(r-r_+)(r-r_-)}dr\,.
\ee
The large-$r$ behavior of the solution in the near region is obtained with the change of variable $z
\rightarrow 1-z$ in the hypergeometric function \cite{abramowitz}. The result is:
\begin{widetext}
\beq
R &\sim &
\left(\frac{r}{r_+-r_-}\right)^{l}\frac{\Gamma(2l+1)}{\Gamma(l+1)}
\left[A\,\frac{\Gamma(1 - 2i\varpi)}{\Gamma(1 + l - 2i\varpi)} +
B\,\frac{\Gamma(1 + 2i\varpi)}{\Gamma(l+1+2i\varpi)}\right]+\nonumber\\
&&
+\left(\frac{r}{r_+-r_-}\right)^{-l-1}\frac{\Gamma(-1-2l)}{\Gamma(-l)}
\left[A\,\frac{\Gamma(1-2i\varpi)} {\Gamma(-l-2i\varpi)}+
B\,\frac{\Gamma(1+2i\varpi)}{\Gamma(-l+2i\varpi)}\right]\,.
\label{near field-large r}
\eeq
\end{widetext}
The matching of the near- and far-region solutions in the region $M \ll r-r_+ \ll 1/\omega$
yields
\be
\frac{B}{A}=-\frac{R_{1}+i(-1)^{l}{(\omega(r_{+}-r_{-}))}^{2l+1}L
R_{3}}{R_{2}+i(-1)^{l} {(\omega(r_{+}-r_{-}))}^{2l+1}L
R_{4}}\,,\label{relation matching B A}\ee
where
\be
L=\frac{\pi}{2^{2l+2}}\frac{{(\Gamma(l+1))}^{2}}{\Gamma(l+3/2)\Gamma(2l+2)\Gamma(2l+1)
\Gamma(l+1/2)}
\ee
and
\be
R_{1}=\displaystyle\frac{\Gamma(1 - 2i\varpi)}{\Gamma(1 + l - 2i\varpi)}\,,\quad
R_{2}=\frac{\Gamma(1 + 2i\varpi)}{\Gamma(l + 1 + 2i\varpi)}\,,\quad
R_{3}=\displaystyle\frac{\Gamma(1-2i\varpi)}{\Gamma(-l-2i\varpi)}\,,\quad
R_{4}=\frac{\Gamma(1+2i\varpi)}{\Gamma(-l+2i\varpi)}\,.
\ee
Equation (\ref{relation matching B A}) can be rewritten as
\begin{widetext}
\be
\frac{B}{A}=-\prod_{k=1}^{l}\left
(\frac{k+2i{\varpi}}{k-2i{\varpi}}\right) \left[
\frac{1+2L{(r_{+}-r_{-})}^{2l+1}\varpi\,\omega^{2l+1}\prod_{k=1}^l
(k^2+4\varpi^2)}{1-2L{(r_{+}-r_{-})}^{2l+1}\varpi\,\omega^{2l+1}\prod_{k=1}^l
(k^2+4\varpi^2)}\right]\,,
\label{relation matching final}
\ee
\end{widetext}
where
\be
\frac{R_{1}}{R_{2}}=\prod_{k=1}^{l}\left(\frac{k+2i{\varpi}}{k-2i{\varpi}}\right)\,,
\nonumber
\ee
\be
R_{3}=iR_{1}(-1)^{l+1}2\varpi\prod_{k=1}^{l}(k^{2}+4{\varpi}^{2})\,,
\nonumber
\ee
and similar relations for $R_{4}$ and $R_{2}$ with $\varpi\rightarrow -\varpi$ have been used. If the
mirror is located near the outer horizon at a radius $r=r_0$, the scalar field must vanish at the mirror
surface. This condition implies
\be
\frac{B}{A}{z_{0}}^{2i\varpi}=-\frac{F(l+1,l+1-2i\varpi,1-2i\varpi,z_{0})}
{F(l+1,l+1+2i\varpi,1+2i\varpi,z_{0})}\,,
\label{relation mirror no approx}
\ee
where $z_{0}=z(r_{0})$. The relation between the position of the mirror and the frequency of the scalar
wave is obtained from Eq.~(\ref{relation matching final}):
\be
\frac{F(l+1,l+1-2i\varpi,1-2i\varpi,z_{0})}{F(l+1,l+1+2i\varpi,1+2i\varpi,z_{0})}={z_{0}}^{2i\varpi}
\prod_{k=1}^{l}\left (\frac{k+2i{\varpi}}{k-2i{\varpi}}\right)
\left[\frac{1+2L{(r_{+}-r_{-})}^{2l+1}\varpi\,\omega^{2l+1}\prod_{k=1}^l(k^2+4\varpi^2)}
{1-2L{(r_{+}-r_{-})}^{2l+1}\varpi\,\omega^{2l+1}\prod_{k=1}^l(k^2+4\varpi^2)}\right]\,.
\label{relation mirror zeros}
\ee
In general, Eq.~(\ref{relation mirror zeros}) must be solved numerically. However, an approximate solution
can be obtained by assuming ${\rm Re}(\omega)\gg {\rm Im}(\omega)$ and $\varpi\ll1$, i.e.\ a frequency near
the superradiant limit $\omega\approx m\Omega$. This solution gives a good approximation for $M\omega\ll1$,
small $m$ and slowly rotating objects. Since ${\rm Re}(\omega)/{\rm Im }(\omega)\ll1$, Eq.~(\ref{relation
mirror zeros}) can be first solved for real $\omega$, then a small imaginary part is added and the equation
is solved again for ${\rm Im}(\omega)$. The l.h.s.\ and the last two terms of the r.h.s.\ of
Eq.~(\ref{relation mirror zeros}) are $\sim 1$ for frequencies near the superradiant limit. This yields
${z_{0}}^{2i\varpi}=1$. Using the tortoise coordinate, it follows
\be
e^{2i\varpi r_{*}^{0}(r_{+}-r_{-})/(2Mr_+)}=e^{2i\varpi
x}=1\,,\label{relation first zeros}
\ee
where $r_{*}^{0}=r_{*}(z_{0})$ and $x=r_{*}^{0}(r_{+}-r_{-})/(2Mr_+)=\log(z_{0})$. The
solution of Eq.~(\ref{relation first zeros}) is
\be
\omega_{n,m}=\frac{\pi(r_{+}-r_{-})}{r_{+}^{2}x}n+m\Omega\,.
\label{relation solution real}
\ee
Positive frequencies can be obtained by imposing $x<-n\pi(r_{+}-r_{-})/(m\Omega{r_{+}^{2}})$. The
superradiant limit requires $\varpi=n\pi(r_{+}-r_{-})/(r_{+}^{2}x)\ll 1$. This condition is satisfied by
considering only the fundamental tone and the first overtones or placing the mirror very close to the
horizon, $|x|\gg 1$. By adding a small imaginary part to the resonant frequency, $\omega =
\omega_{n,m}+i\delta$, where $\delta<<\omega_{n,m}$, Eq.~(\ref{relation mirror zeros}) becomes
\begin{widetext}
\beq &
&\frac{F(l+1,l+1-2i\left(\varpi_{0}+i\rho\delta\right),1-2i\left(\varpi_{0}+i\rho\delta\right),z_{0})}
{F(l+1,l+1+2i\left(\varpi_{0}+i\rho\delta\right),1+2i\left(\varpi_{0}+i\rho\delta\right),z_{0})}
{z_{0}}^{-2i\varpi_{0}}{z_{0}}^{2\rho\delta}=\nonumber \\
 & & =\prod_{k=1}^{l}\left (\frac{k+2i\varpi_{0}}{k-2i\varpi_{0}}\right)\left[\frac{1+2L{(r_{+}-
 r_{-})}^{2l+1}\varpi_{0}\, \omega^{2l+1}_{n,m}\prod_{k=1}^l (k^2+4{\varpi_{0}}^2)}
 {1-2L{(r_{+}-r_{-})}^{2l+1}\varpi_{0}\,\omega^{2l+1}_{n,m}
\prod_{k=1}^l (k^2+4{\varpi_{0}}^2)}\right]\,,
\label{relation mirror zeros complete}
\eeq
\end{widetext}
where $\varpi=\varpi_{0}+i\rho\delta$ and $\rho=r_{+}^{2}/(r_{+}-r_{-})$. The ratio of the hypergeometric
functions in the l.h.s.\ of Eq.~(\ref{relation mirror zeros complete}) is $\sim 1$ for the approximations
used in the derivation. In terms of the tortoise coordinate, Eq.~(\ref{relation mirror zeros complete})
reads
\be
e^{2\rho
x\delta}=\frac{1+2L{(r_{+}-r_{-})}^{2l+1}\varpi_{0}\,\omega^{2l+1}_{n,m}
\prod_{k=1}^l (k^2+4{\varpi_{0}}^2)}
{1-2L{(r_{+}-r_{-})}^{2l+1}\varpi_{0}\,\omega^{2l+1}_{n,m}
\prod_{k=1}^l (k^2+4{\varpi_{0}}^2)}\,.
\label{e2rho}
\ee
The solution of Eq.~(\ref{e2rho}) is
\be
\delta={\rm Im}(\omega)=\frac{r_{+}-r_{-}}{2r_{+}^{2}x}\log\left[\frac{1+\gamma(\omega_{n,m}-m\Omega)}
{1-\gamma(\omega_{n,m}-m\Omega)}\right]\,,
\label{relation imaginary first solution}
\ee
where
\be \gamma\equiv \frac{\pi}{2} r_{+}^{2}{\left
(\frac{r_{+}-r_{-}}{2}\right )}^{2l}\prod_{k=1}^l
(k^2+4{\varpi_{0}}^2)
\frac{[\Gamma(l+1)]^{2}}{\Gamma(l+3/2)\Gamma(2l+2)\Gamma(2l+1)\Gamma(l+1/2)}
\omega^{2l+1}_{n,m} \geq 0\,.
\label{relation constant gamma}
\ee
Both $\delta$ and $\omega_{n,m}$ are very small for $r_{0}\sim r_{+}$. However, the argument of the
logarithm in Eq.~(\ref{relation imaginary first solution}) is $\sim 1$ and the assumption ${\rm
Re}(\omega)\gg {\rm Im}(\omega)$ is satisfied.

The above results display two important features. First, Eq.~(\ref{relation imaginary first solution}) and
Eq.~(\ref{relation constant gamma}) imply $\delta\lessgtr 0$ for ${\rm Re}(\omega)\gtrless m\Omega$. The
time dependence of the scalar field $\Phi$ is $\exp(-i\omega t)= \exp(-i {\rm Re}(\omega) t)\exp(\delta
t)$. Thus the amplitude of the field grows exponentially and the resonant mode becomes instable for ${\rm
Re}(\omega)<m\Omega$. Second, $x<x_{\rm crit}=-n\pi(r_{+}-r_{-})/(m\Omega{r_{+}^{2}})$ and ${\rm
Re}(\varpi_{0})\propto n x^{-1}$ imply $\delta\lessgtr 0$ for $n\lessgtr 0$. There is always a superradiant
amplification for $n>0$ provided that the mirror position is closer than $x_{crit}$ to the horizon and the
approximations used in the above derivation are satisfied. The critical value $x_{\rm crit}$ is positive
and outside the domain of $x$ for $n<0$. In this case the mirror can be located everywhere in the
near region, but there is no superradiant amplification. The growth timescale for $n>0$ is given by
\be
\tau\equiv\delta^{-1}=2\rho x\log^{-1}\left[\frac{1+\gamma(\omega_{n,m}-m\Omega)}
{1-\gamma(\omega_{n,m}-m\Omega)}\right]\,,
\label{relation time scale}
\ee
or, in terms of the physical variables, by Eq.~(\ref{relation time scale unit2}).
\subsection{\label{app:scalar mass} Massive scalar field}
If the scalar field is massive, the wave equation is
\be
\Delta_{\mu}\Delta^{\mu}\Psi=\mu^{2}\Psi\,,
\ee
where $\Delta_\mu$ the covariant derivative and $\mu$ is the field mass. The above equation is
separable. The radial equation is
\be \Delta\frac{d}{dr}\left(\Delta\frac{dR}{dr}\right)+
[\omega^{2}(r^{2}+a^{2})^{2}- 4aM\omega r+
+a^{2}m^{2}-\Delta(\mu{2}r^{2}+a^{2}\omega^{2}+\lambda)]R=0\,.
\label{wave eq separated mass}
\ee
Assuming $a\omega\ll 1$, $\lambda\approx l(l+1)$ and $\mu r\ll l$, the solution in the near region is
identical to the solution for the massless case. The equation in the far region is
\be
\partial_r^2 (r R)+ \left [ \omega^2-\mu^{2}-l(l+1)/r^2
\right ] (r R)=0\,.
\ee
The results of the massless case apply with the substitution $\omega\rightarrow k$, where
$k^{2}=\omega^{2}-\mu^{2}>0$. The matching conditions are
\be
\frac{F(l+1,l+1-2i\varpi,1-2i\varpi,z_{0})}{F(l+1,l+1+2i\varpi,1+2i\varpi,z_{0})}={z_{0}}^{2i\varpi}
\prod_{k=1}^{l}\left (\frac{k+2i{\varpi}}{k-2i{\varpi}}\right)
\left[\frac{1+2L{(r_{+}-r_{-})}^{2l+1}\varpi
{(\omega^{2}-\mu^{2})}^{l+1/2} \prod_{k=1}^l
(k^2+4\varpi^2)}{1-2L{(r_{+}-r_{-})}^{2l+1}\varpi
{(\omega^{2}-\mu^{2})}^{l+1/2}\prod_{k=1}^l (k^2+4\varpi^2)}\right].\nonumber
\ee
The imaginary part of the frequency is identical to Eq.~(\ref{relation imaginary first
solution}) with
\be
\gamma\equiv \pi r_{+}^{2}{\left (\frac{r_{+}-r_{-}}{2}\right
)}^{2l}\prod_{k=1}^l (k^2+4{\varpi_{0}}^2)
\left[\frac{[\Gamma(l+1)]^{2}}{\Gamma(l+3/2)\Gamma(2l+2)\Gamma(2l+1)\Gamma(l+1/2)}\right]
(\omega^{2}_{n,m}-\mu^{2})^{l+1/2}\geq 0\,.
\ee
The condition for superradiant amplification, $\omega <m\Omega$, does not depend on the field
mass.
\subsection{\label{app:general spin} General spin-$s$ fields}
Previous analytical and numerical calculations \cite{staro1,Teukolsky:1974yv} have shown that superradiant
effects for gravitational fields are stronger  than for scalar fields. For instance, superradiant
amplification factors are about 0.1\%, 4.5\% and 138\% for scalar, electromagnetic and gravitational field,
respectively. Since the effects induced by the BH in the far region can be neglected, the radial wave
equation reduces to the wave equation of a massless field with spin-weight $s$, frequency $\omega$ and
angular momentum $l$ in flat background:
\be
r\partial_r^2 f+ 2(1+l-i\omega r)\partial_r f-2i(1+l-s)\omega
f=0\,,
\ee
where $R=e^{-i\omega r}r^{l-s}\,f(r)$. Introducing the radial coordinate $x=2i\omega r$, the
wave equation becomes a standard Kummer equation \cite{abramowitz}. Its most general solution
is a linear combination of confluent hypergeometric functions:
\be
R =e^{-i\omega r} r^{l-s} (\alpha\,M(1+l-s,2l+2,2i\omega
r)+\beta\,U(1+l-s,2l+2,2i\omega r)) \,.
\label{confluent solution general}
\ee
The large-$r$ behavior of  Eq.~(\ref{confluent solution general}) is
\be
R\sim\left[\alpha\frac{\Gamma(2l+2)}{\Gamma(1+l-s)}(2i\omega)^{-1-l-s}
\right]\frac{e^{i\omega r}}{r^{1+2s}}+
\left[\alpha(-1)^{1+l-s}\frac{\Gamma(2l+2)}{\Gamma(1+l+s)}+\beta\right]
\frac{e^{-i\omega r}}{r}\,.
\label{far field-large r gen}
\ee
The first two terms in Eq.~(\ref{far field-large r gen}) represent an outgoing wave at infinity
and an incoming wave from infinity, respectively. The behavior for small $r$ is
\be
R\sim\left[\alpha+\beta\frac{\Gamma(-1-2l)}{\Gamma(-l-s)}\right]\,
r^{l-s}+\beta(2i\omega)^{-1-2l}\frac{\Gamma(2l+1)}{\Gamma(1+l-s)}r^{-1-l-s}\,.
\label{far field-small r gen}
\ee
Absence of ingoing waves at infinity implies
\be
\beta=-\alpha(-1)^{1+l-s}\frac{\Gamma(2l+2)}{\Gamma(1+l+s)}\,.
\label{relation alpha beta gen}
\ee
The near-region behavior of the solution in the far region is
\be R \sim  \alpha \left[ \frac{1}{2}\:r^{l-s} -
(2i\omega)^{-1-2l}(-1)^{1+l-s}\frac{\Gamma(2l+2) \Gamma(2l+1)}{\Gamma(1+l-s)\Gamma(1+l+s)}\:
r^{-l-1-s}\right]\,.
 \label{far field-small r2u gen}
\ee
The radial wave equation in the near region is
\beq
\Delta^{-s}\frac{d}{dr}\left ( \Delta^{s+1}\frac{dR}{dr}\right
)+\left[\frac{K^{2}-2is(r-M)K}{\Delta}- \lambda \right]R=0\,
,\label{near eq gen}\nonumber
\eeq
where $\lambda=(l-s)(l+s+1)+O(\omega a)$. Using the approximate relations
\be
\Delta'\approx r_+-r_-\,,\quad
K=\omega(r^{2}+a^{2})-am\approx
r_{+}^{2}(\omega-m\Omega)\,,\quad
K^{2}-2is(r-M)K\approx (r_+-r_-)^{2}\left[\frac{s^2}{4}+\left(\varpi
-i\frac{s}{2}\right)^2\right]\,,\nonumber
\ee
and introducing the radial coordinate $z$, Eq.~(\ref{near eq gen}) can be written as
\be
z(1-z)\partial_z^2 R+ ((s+1)+(s-1)z)\partial_z R +\left\{\frac{1-z}{z}\left[\frac{s^2}{4}+(\varpi -is/2)^2
\right]-\frac{(l-s)(l+s+1)}{1-z}\right\} R=0\,.
\ee
Setting $R=z^{i \,\varpi} (1-z)^{l+s+1}\,F$, the previous equation becomes a
standard hypergeometric equation \cite{abramowitz}. Its most general solution is
\be
R =A\, z^{-s-i\,\varpi}(1-z)^{l+1+s} F(a-c+1,b-c+1,2-c,z)+B\,z^{i\,\varpi}(1-z)^{l+1+s} F(a,b,c,z) \,,
\label{hypergeometric solution general}
\ee
where $a=1+l+s+i\,2\varpi$, $b=l+1$ and $c=1+s+ i\,2\varpi$. The behavior of
Eq.~(\ref{hypergeometric solution general}) near $r_+$ is
\be
R \sim A\,e^{-s-i(\omega-m\Omega)\frac{r_+}{2M}r_*}+
B\,e^{i(\omega-m\Omega)\frac{r_+}{2M}r_*}\,.
\ee
The large-$r$ behavior is
\be R \sim
\left(\frac{r}{r_+-r_-}\right)^{l-s}\frac{\Gamma(2l+1)}{\Gamma(l+1)}{(A\,R_1+ B\,R_2)}+
\left(\frac{r}{r_+-r_-}\right)^{-l-1-s}\frac{\Gamma(-1-2l)}{\Gamma(-l)}{(A\,R_3+B\,R_4)}\,,
\label{near field-large r gen}
\ee
where
\beq R_1
&=&\frac{\Gamma(1-s-2i\varpi)}{\Gamma(l-s+1-2i\varpi)}\,,\qquad R_2
=\frac{\Gamma(1+s+2i\varpi)}
{\Gamma(l+s+1+2i\varpi)}\,,\nonumber \\
R_3 &=&\frac{\Gamma(1-s-2i\varpi)}{\Gamma(-l-s-2i\varpi)}=R_1(-1)^{l+1}(s+2i\varpi)\prod_{k=1}^{l}
(k^2+4\varpi^{2}-s^{2}-4is\varpi)\,,\nonumber \\
R_4 &=&\frac{\Gamma(1+s+2i\varpi)}{\Gamma(-l+s+2i\varpi)}=-R_2(-1)^{l+1}(s+2i\varpi)\prod_{k=1}^{l}
(k^2+4\varpi^{2}-s^{2}-4is\varpi)\,.
\label{relation coeff gen}\nonumber\\
\eeq
The matching of Eq.~(\ref{near field-large r gen}) and Eq.~(\ref{far field-small r2u gen})
yields
\be
\frac{B}{A}=-\prod_{k=1}^{l}\left(\frac{k+s+2i\varpi}{k-s-2i\varpi}\right)\frac{1-i\gamma
L_{s}(s+2i\varpi) g_s}{1+i\gamma
L_{s}(s+2i\varpi)g_s}\,,\label{relation matching final gen}\ee
where $\gamma=(w(r_+ -r_-))^{2l+1}$, $g_s=\prod_{k=1}^{l}(k^2+4\varpi^{2}-s^{2}-4is\varpi)$ and
$L_s$ is defined in Eq.~(\ref{Ls})  for $l \neq \pm s-1$. Equation (\ref{relation matching
final gen}) reduces to Eq.~(\ref{relation matching final}) for $s=0$.

If the mirror is close to the outer horizon, the field must vanish at the mirror
surface. Using the small-$r$ behavior of the solution in the near region and setting
the radial field to zero at $z=z_{0}$, it follows $A=-Bz_{0}^{s+2i\varpi}$. The
relation between the position of the mirror and the frequency of the wave is given by
Eq.~(\ref{relation mirror zeros gen2}).
\section{\label{app:extremal} Analytic solution in the near-extremal regime}
This appendix describes near-extremal horizonless objects by employing the analytic approximations of
Refs.~\cite{staro1,Teukolsky:1974yv}. The resonant frequencies for rapidly spinning BHs can be found by
rewriting the Teukolsky equation in the Kerr incoming coordinate system ($v$,$r$,$\theta$,${\phi'}$). The
latter is obtained from the Boyer-Lindquist coordinates with the coordinate transformation
$dv=dt+(r^{2}+a^{2})dr/\Delta$, $d{\phi'}=d\phi+adr/\Delta$. The radial equation for a field of spin $s$ is
\cite{Teukolsky:1974yv}
\be
\Delta R''(r)+2[(s+1)(r-M)-iK]R'(r)-\left[2(2s+1)i\omega r-\lambda\right]R(r)=0\,.
\label{radial extremal}
\ee
Defining $x=(r-r_+)/r_+$, $\sigma=(r_+-r_-)/r_+$, $\tau=M(\omega-m\Omega)$ and
$\omega'\rightarrow r_{+} \omega$, Eq.~(\ref{radial extremal}) reads
\be
x(x+\sigma)
R''(x)-\left[2(2s+1)i\omega'(x+1)+\lambda\right]R(x)-\left[2i\omega'
x^2+x(4i\omega'-2(s+1))-(s+1)\sigma++4i\tau\right]R'(x)=0\,.
\label{radial extremal redef}
\ee
In the far region, $x\gg \max(\sigma,\tau)$, Eq.~(\ref{radial extremal redef})
reduces to
\be x^{2} R''(x)-[2i\omega' x^2+x(4i\omega'-2(s+1))]R'(x)-[2(2s+1)i\omega'(x+1)+\lambda]R(x)=0\,.
\label{radial extremal far}
\ee
Its most general solution is written in terms of confluent hypergeometric functions
\be R(x)=\alpha x^{-s-\frac{1}{2}+2i\omega'+i\delta}M\left(\frac{1}{2}+s+2i\omega'+i\delta,1+2i\delta,2i\omega x\right)
+ \beta\,(\delta\rightarrow -\delta)\,,
\label{extremal far solution}
\ee
where $\delta^2=4\omega'^{2}-(s+1/2)^2-\lambda$. The large-$x$ behavior of Eq.~(\ref{extremal far
solution}) is
\be
R\sim x^{-1-2s}\left[\frac{\alpha\Gamma(1+2i\delta)
(e^{i\pi}/2i\omega')^{1/2+s+2i\omega'+i\delta}
}{\Gamma(1/2-s-2i\omega'+i\delta)}\right]+\frac{e^{2i\omega'x}}{x^{1-4i\omega'}}
\left[\frac{\alpha\Gamma(1+2i\delta)(2i\omega')^{-1/2+s+2i\omega'-i\delta}
}{\Gamma(1/2+s+2i\omega'+i\delta)}\right]+\beta\,(\delta\rightarrow-\delta)\,.
\label{extremal far large sol}
\ee
The term representing an incoming (outgoing) wave at infinity behaves as $r^{-1-2s}=e^{2i\omega r_*}/r$ for
near-extremal BHs. (See Table 1 of Ref.~\cite{Teukolsky:1974yv}). Thus the first and second terms in
Eq.~(\ref{extremal far large sol}) describe an incoming wave and an outcoming wave, respectively. Absence
of incoming waves at infinity implies
\be
\beta=-\alpha\frac{\Gamma(1+2i\delta)}{\Gamma(1-2i\delta)}\frac{\Gamma(1/2-s-2i\omega'-i\delta)}
{\Gamma(1/2-s-2i\omega'+i\delta)}\left(
{-2i\omega'}\right)^{-2i\delta}\,.\label{rel beta alpha extremal}
\ee
Equation (\ref{extremal far solution}) reduces to
\begin{eqnarray}
R \sim \alpha\, x^{-s-1/2+2i\omega'+i\delta}+\beta\,x^{-s-1/2+2i\omega'-i\delta}
\label{extremal far small sol}
\end{eqnarray}
for small $r$. In the near region, Eq.~(\ref{radial extremal redef}) reads
\be
x(x+\sigma) R''(x)-[x(4i\omega'-2(s+1))-(s+1)\sigma+4i\tau]R'(x)+[2(2s+1)i\omega'+\lambda]R(x)=0\,.
\nonumber
\ee
The radial wave equation in the near region becomes a standard hypergeometric equation with the
substitution $z=-x/\sigma$. Its most general solution is
\be
R=A\,z^{1-c}F(a-c+1,b-c+1,2-c,z)+B\,F(a,b,c,z)\,,
\ee
where $a=1/2+s-2i\omega'+i\delta$, $b=1/2+s-2i\omega'-i\delta$, $c=1+s+i\kappa$ and $\kappa=-4\tau/\sigma$.
The mirror is located at $r\sim r_+$. In this limit the solution behaves as $R \sim A\, z^{1-c}+B$. The
large-$r$ behavior of the solution in the near region is
\beq
&&R\sim
\left(\frac{x}{\sigma}\right)^{-a}\left[A\,(-1)^{1-c}\frac{\Gamma(2-c)\Gamma(b-a)}{\Gamma(b-c+1)
\Gamma(1-a)}+B\,\frac{\Gamma(c)\Gamma(b-a)}{\Gamma(b)\Gamma(c-a)}
\right]+\\ \nonumber
&&~\hskip 80pt+\left(\frac{x}{\sigma}\right)^{-b}\left[A\,(-1)^{1-c}\frac{\Gamma(2-c)\Gamma(a-b)}{\Gamma(a-c+1)
\Gamma(1-b)}+B\,\frac{\Gamma(c)\Gamma(a-b)}{\Gamma(a)\Gamma(c-b)}\right]\,,
\label{near field-large r extremal}
\eeq
or, in terms of the physical variables,
\be
R\sim \left(\frac{x}{\sigma}\right)^{-1/2-s+2i\omega'+i\delta}
\Gamma(2i\delta)\left[A\,(-1)^{s+i\kappa}R_1+B\,R_2\right]+(\delta\rightarrow -\delta)\,,
\label{extremal near large sol}
\ee
where
\beq
&&R_1(\delta)=\frac{\Gamma(1-s-i\kappa)}{\Gamma(1/2-s+2i\omega'+i\delta)
\Gamma(1/2-2i\omega'+i\delta-i\kappa)}\,,\nonumber\\
&&R_2(\delta)=\frac{\Gamma(1+s+i\kappa)}{\Gamma(1/2+s-2i\omega'+i\delta)
\Gamma(1/2+2i\omega'+i\delta+i\kappa)}\,,\nonumber\\
&&R_3=R_1(-\delta)\,,\qquad
R_4=R_2(-\delta)\,.
\label{relation coeff extremal}
\eeq
The matching in the overlapping region yields
\be
\alpha=\sigma^b \Gamma{(a-b)}\left[A\,(-1)^{1-c}R_1+B\,R_2\right]\,,\qquad
\beta=\sigma^a \Gamma{(b-a)}\left[A\,(-1)^{1-c}R_3+B\,R_4\right]\,.
\label{relation matching extremal}
\ee
Using Eq.~(\ref{rel beta alpha extremal}), it follows
\be
-\rho(-2i\omega'\sigma)^{2i\delta}=\frac{A(-1)^{s+i\kappa}\,R_1+B\,R_2}{A(-1)^{s+i\kappa}\,R_3+B\,R_4}\,,
\label{matching extremal check}
\ee
where
\be
\rho=\frac{\Gamma(-2i\delta)\Gamma(1-2i\delta)\Gamma(1/2-s-2i\omega'+i\delta)}{\Gamma(2i\delta)
\Gamma(1+2i\delta)\Gamma(1/2-s-2i\omega'-i\delta)}\,.
\label{relation matching final extremal}
\ee
Equation (\ref{matching extremal check}) reduces to Eq.~(9) of Ref.~\cite{Detweiler:1980gk} for suitable
boundary condition in presence of a horizon, i.e.\ $A=0$. In the case under study, $A$ is nonvanishing
because the region containing the horizon is forbidden by the presence of the mirror. Thus
Eq.~(\ref{relation matching extremal}) yields
\be
\frac{B}{A}=-(-1)^{s+i\kappa}\frac{R_{1}+\rho
R_{3}(-2i\sigma\omega')^{2i\delta}}{R_{2}+\rho
R_{4}(-2i\sigma\omega')^{2i\delta}}\,.
\label{relation matching B A extremal}
\ee
Dirichlet boundary conditions at the wall require $z_{0}^{-s-i\kappa}=-B/A$. The relation between the
position of the mirror and the frequency of the wave, Eq.~(\ref{extremal zero equation2}), is obtained from
Eq.~(\ref{relation matching final extremal}).

\end{document}